%% file: main.tex
\documentclass[]{aastex631}

\usepackage{bm}

\shorttitle{Gamma-ray and Neutrino Coincidences}
\shortauthors{Ayala Solares et al.}
\graphicspath{{./}{figures/}}

\begin{document}

\title{Search for Gamma-Ray and Neutrino Coincidences Using HAWC and ANTARES Data}

\correspondingauthor{Hugo Alberto Ayala Solares}
\email{hgayala@psu.edu}

\input {authors}



\begin{abstract}
In the quest for high-energy neutrino sources, the Astrophysical Multimessenger Observatory Network (AMON) has implemented a new search by combining data from the High Altitude Water Cherenkov (HAWC) observatory and the Astronomy with a Neutrino Telescope and Abyss environmental RESearch (ANTARES) neutrino telescope. Using the same analysis strategy as in a previous detector combination of HAWC and IceCube data, we perform a search for coincidences in HAWC and ANTARES events that are below the threshold for sending public alerts in each individual detector. Data were collected between July 2015 and February 2020 with a livetime of 4.39 years. Over this time period, 3 coincident events with an estimated false-alarm rate of $<$ 1 coincidence per year were found. This number is consistent with background expectations. 
\end{abstract}



\section{Introduction} \label{sec:intro}

The Astrophysical Multimessenger Observatory Network \citep[AMON][]{amon2020} is a virtual hub that integrates heterogeneous data from different astrophysical observatories with the main objective of enabling multimessenger astrophysics. Observatories that become members of AMON can act as trigger observatories or as follow-up observatories. Triggering observatories have high-duty cycles and a large field of view. Follow-up observatories have better angular resolution and sensitivity. AMON has developed coincidence analyses between high-energy gamma-ray and high-energy neutrino data. AMON mainly, but not necessarily, receives and uses data that are below the astrophysical event selection threshold (called subthreshold) for the individual observatories. In these data, possible signal events of astrophysical origin can be present and due to the limited sensitivity of a given detector (e.g., HAWC or ANTARES), cannot be distinguished from background events. Using a statistical analysis, AMON looks for temporal and/or spatial coincidences between events collected by different observatories with the purpose of recovering the signal events that are buried in the background.

The AMON analyses using gamma-ray and neutrino data include the coincidence studies between IceCube and \textit{Fermi}-LAT  \citep{amonfermi_icecube}; ANTARES and \textit{Fermi}-LAT\citep{amonfermi_antares}; and HAWC and IceCube \citep{amonHAWCIC}. The last two analyses make use of the Neutrino-Electromagnetic (NuEM) AMON channel. This channel generates alerts in real time after receiving data from the respective observatories and performing a calculation to rank the coincidences (see Section \S\ref{sec:analysis}). AMON servers are now located at the Amazon Web Services (AWS), having a high up-time ($>$99.99\%). The NuEM alerts are sent as notices and circulars to the \textit{Gamma-ray Coordinates Network} \citep[GCN;][]{gcn}. Recently, AMON also started to send alerts to the \textit{Scalable Cyberinfrastructure to support Multi-Messenger Astrophysics} \citep[SCiMMA;][]{scimma,scimma2}, a new hub for multimessenger astrophysics designed for private and public communication.

The NuEM channel searches for sources that emit secondary neutrinos and gamma rays. These neutrinos and gamma rays are produced in hadronic interactions, such as inelastic collisions of cosmic rays with matter or with radiation fields. These hadronic interactions produce neutral and charged pions, which then decay into the aforementioned particles.  These interactions can occur in a wide variety of sources such as blazar flares, tidal disruption events, long gamma-ray bursts, short gamma-ray bursts, supernovae, and compact binary mergers \citep[for a review of multimessenger sources, see][]{mmsources}.
In this work, we present a new analysis of this channel: the coincidence search between events collected by the HAWC gamma-ray observatory and the ANTARES neutrino telescope.

With ANTARES recently ceasing operations, this analysis helps us not only to look for possible sources in existing data, but also to prepare the necessary analysis tools and infrastructure for the KM3NeT neutrino telescope \citep{km3net}, the successor of ANTARES. 

\section{HAWC and ANTARES: Detectors and Data sets}

\subsection{HAWC}
The High Altitude Water Cherenkov (HAWC) observatory monitors the gamma-ray sky from its location in Puebla, Mexico, at an altitude of 4100 meters above sea level. Sitting between the volcanoes Sierra Negra and Pico de Orizaba, it has a large field of view that covers two-thirds of the sky daily. 
With a duty cycle above $95\%$, HAWC can monitor 2 sr of the sky continuously, which makes it ideal for observing transient events \citep{hawc}. 
HAWC is a water Cherenkov detector array that characterizes the footprints of extensive air showers. 
Hadron-like showers and gamma-like showers can be classified by looking at how smooth and compact is the distribution of the charge measured by the photomultiplier tubes (PMTs) in the array.   
Hadron-like showers tend to have a discontinuous profile on the array due to the large number of muons in the shower, while gamma-ray showers present a smoother profile. By using the trigger time information of each PMT, reconstruction algorithms can find the direction of the primary particle with a 68\% resolution of $\sim 0.2^{\circ}$ at energies above 10 TeV. HAWC is sensitive to gamma rays with energy from 300\,GeV up to $>$100\,TeV \citep{hawc, 3hwc}. 

The data that AMON receives from HAWC for this analysis includes the rising and setting time of the event position in the sky with respect to the detector ---which defines the ``HAWC transit time'' of the event; a parameter (referred to as the ``significance value'' in the following) that estimates how much the event deviates from the expected hadron-like background and it is calculated after one transit; and the position in the sky of every event with their uncertainty. HAWC events are referred as ``hotspots''. The data used in this work were collected from July 2015 to February 2020.

\subsection{ANTARES}
The ANTARES neutrino telescope \citep{antaresDet} is located 40 km off-shore from the city of Toulon, France, in the Mediterranean Sea. It is a deep-sea Cherenkov neutrino detector. The detector consists of a three-dimensional array of 885 optical modules, each one with a 10~inch PMT, and distributed over 12 vertical strings anchored in the seafloor at a depth of about 2400~m. The detection of light from up-going charged particles is optimized with the PMTs facing 45$^{\circ}$ downward. Since May 2007, the telescope has detected neutrino-induced muons that cause the emission of Cherenkov light in the detector, producing \textit{track-like} events. Charged-current interactions induced by electron neutrinos (and, possibly, by tau neutrinos of cosmic origin) or neutral-current interactions of all neutrino flavors can be reconstructed as \textit{cascade-like} events \citep{antaresDet2}. For the analysis presented in this manuscript, we use track-like events that are used in the point-source search analysis of ANTARES \citep{giuliaICRC}, which have a median angular resolution of $0.4^{\circ}$ for energies above 10 TeV. Since the ANTARES data is public, we are not using subthreshold data from ANTARES for this analysis.

The ANTARES data information consists of the following: the position and uncertainty of the individual observed event, the time of the event, and a $p$-value that quantifies the probability of the event to be a background event. 
For this study, we use the archival public data (2007-2017) that can be found in \citep{antaresPublic} as well as 3 more years of archival data (up to 2020) given by ANTARES through the AMON memorandum of understanding. Since we are using the ANTARES public data, it does not contain subthreshold events. 
We use the data that overlap with the time period of the HAWC data.

\section{The Coincidence Analysis}\label{sec:analysis}

\subsection{Computing the ranking statistics RS}\label{sec:statistic}
The analysis method applied in this work is the same as the one developed for the HAWC and IceCube detectors in \cite{amonHAWCIC}, which is summarized below.

We assume to have a coincidence when the time of the ANTARES event falls between the rising and setting time of the HAWC event and the distance between the reconstructed directions of the events is smaller than 3.5$^{\circ}$ \citep[same as in][]{amonHAWCIC}. 
After finding a coincidence, a test statistic is calculated to rank the coincidence. This is defined as
\begin{equation}\label{eq:chi2}
    \chi^2_{6+2n_{\nu}} = -2 \ln (p_{_{\lambda}} p_{_{\rm HAWC}} p_{_{\rm Cluster}} \prod^{n_{\nu}}_i p_{_{\rm ANTARES,i}}),
\end{equation}
which is based on the Fisher's method \citep{fisher}. The number of degrees of freedom of the test statistic is twice the number of $p$-values.
The $p_{_{\lambda}}$ value measures how much the events spatially overlap with each other. This value is obtained after optimizing a 
log-likelihood function defined as 

\begin{equation}\label{eq:lambda}
\lambda( \bm{x} ) = \sum_{i=1}^N  \ln \frac{S_i(\bm{x})}{B_i}.
\end{equation}

Here $S_i(\bm{x}) =  \exp{[-(\bm{x}-\bm{x}_{i})^2/2\sigma_{i}^2}]/(2 \pi \sigma_{i}^2)$, 
a 2D Gaussian on a sphere with $\bm{x}_{i}$ and $\sigma_{i}$ being the measured position and positional uncertainty of the $i$-th event. $B_i$ is the background directional probability distribution from the corresponding detector at the position of the events. The sum is over all the $N$ events that are part of the coincidence.
The position of the coincidence, $\bm{x}_{\rm coinc}$, is defined as the position where the log-likelihood is maximized, $\lambda_{\rm max}$. The $p_{_{\lambda}}$ is obtained from the $\lambda_{\rm max}$ distribution, which is the probability of seeing a $\lambda_{\rm max}$ or higher. 

The $p_{_{\rm HAWC}}$ value is related to the significance value of the HAWC event and quantifies the probability for the event to be from background. 

The $p_{_{\rm Cluster}}$ is the probability of having $n_{\nu}$ ANTARES events when one is already observed. It is defined as
 \begin{equation}
 p_{_\mathrm{Cluster}(n_{\nu})} = 1 - \sum^{n_{\nu}-2}_{i=0} \mathrm{Poisson}(i;f_{\nu} \Delta t),
 \end{equation} 
 where $\Delta t$ is the HAWC transit time; $f_{\nu}$ is the ANTARES background rate in a 3.5$^{\circ}$ circle in the sky estimated as $f_{\nu} = f_{\rm all}\frac{\Omega}{4\pi} = f_{\rm all}(1-\cos(3.5^{\circ}))/2$, where $f_{\rm all}$ is the measured background rate from the whole sky.

Finally $p_{_{\rm ANTARES,i}}$, is the fraction of ANTARES events that have a larger number of hits in the detector than the observed number of hits for the event. It is computed by using the normalized anti-cumulative distribution of the number of hits from the full ANTARES public data.

Since there can be $n_{\nu}$ ANTARES events passing the selection criteria during a HAWC time window, the degrees of freedom of Eq. \ref{eq:chi2} vary. Therefore, we compute the $p$-value of the $\chi^2_{6+2n_{\nu}}$ with ${6+2n_{\nu}}$ degrees of freedom. The ranking statistic (RS) is then simply defined as 
\begin{equation}\label{eq:rankStat}
    {\rm RS} = -\log_{10}(p{\rm-value}_{\chi^2_{6+2n_{\nu}}}). 
\end{equation}

\subsection{Calculating the \textit{False Alarm Rate}}\label{sec:far}
\begin{figure}
\centering
\includegraphics[scale=0.5]{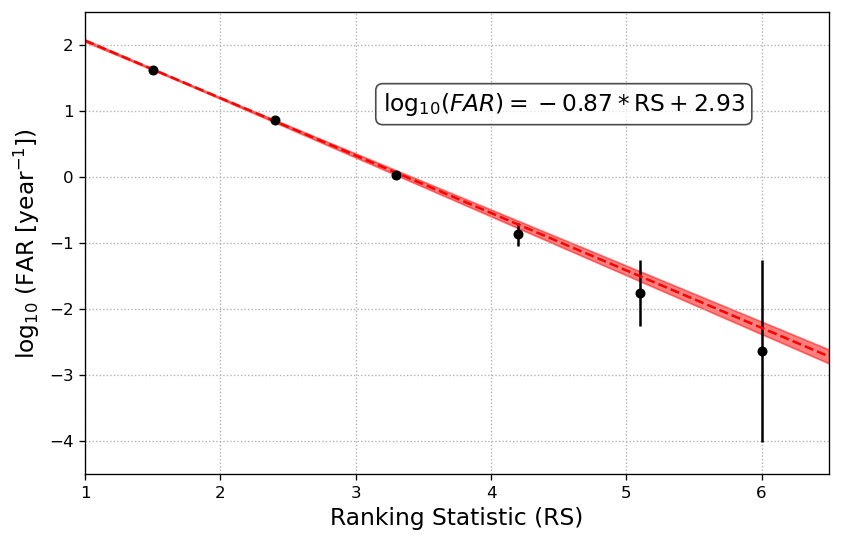}
\caption{False alarm rate (FAR) of the analysis. The result and equation of the linear fit, together with the 1$\sigma$ statistical band, are also shown.}
\label{fig:far}
\end{figure}

The distribution of the RS is used to quantify the probability that the coincidences are fortuitous. It is also used to calculate the false alarm rate (FAR) of the coincidence (i.e. how rare the RS is). 
To build the distribution, we perform several simulations by scrambling the data sets a thousand times. The scrambling consists of randomizing the right ascension and the time values of the events.
We then count how many events are above an RS value and divide by the simulated time ($\sim$4600 years).
Figure \ref{fig:far} shows the FAR as a function of the RS. A red line is shown which fits the data points, together with the 1$\sigma$ uncertainty band.  

\subsection{Sensitivity and Discovery Potential }\label{sec:sensitivity}
\begin{figure}
    \centering
    \includegraphics[scale=0.45]{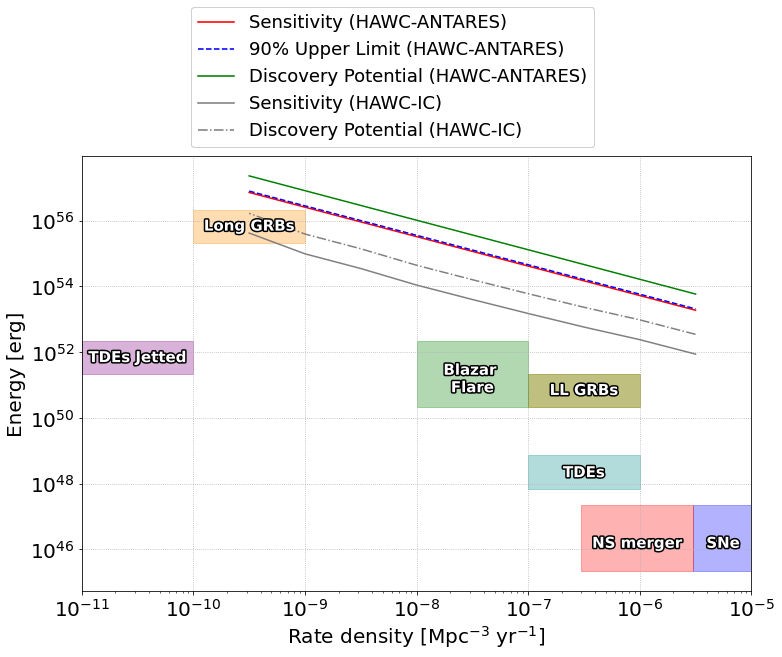}
    \caption{Sensitivity, discovery potential (5$\sigma$) and 90 \% upper limit for the archival data (analysis livetime of 4.39 years) in terms of total isotropic equivalent neutrino energy as a function of the local rate density. We assume a burst time of 6 hours and the neutrino spectrum to be a power law with index $-2.0$. Luminosity and rate-density ranges of the different sources can be found in \cite{mmsources}. For comparison, we show the sensitivity and discovery potential of the HAWC-IceCube analysis from \cite{amonHAWCIC}. Both the ANTARES effective area and the overlap region between HAWC and ANTARES are smaller compared to that in the HAWC-IceCube analysis. We can see that long gamma-ray bursts are potential candidates for a possible coincidence detection.}
    \label{fig:sensitivity}
\end{figure}

We obtain the sensitivity and discovery potential in the parameter space composed of the local rate density of transient sources vs the total neutrino isotropic energy as in \cite{amonHAWCIC}. We compare this to different source populations. 
We use the FIRESONG software package \citep{firesong} to obtain the number of transient sources during the same time as the archival data, along with the redshift, the neutrino flux normalization and the position in the sky of each source. The star formation rate assumed is from the core-collapse supernova rate obtained from the CANDELS and CLASH supernova surveys \citep[][]{ccsnr2015}. We assume a local rate density and a total neutrino isotropic equivalent energy as denoted along the x- and y- axes of Figure \ref{fig:sensitivity}. The duration of each burst is fixed to 6 hours. For the neutrino energy spectrum, we assume a power law with a spectral index of -2.0 in the energy range between 10 TeV and 10 PeV.
We use the model from \cite{prdkohta} given as 
\begin{equation}\label{eq:nuem}
    E_{\gamma} F_{\gamma}(E_{\gamma}) \approx e^{-\frac{d}{\lambda_{\gamma \gamma}}} \frac{2}{3K}\sum_{\nu_{\alpha}} E_{\nu} F_{\nu_{\alpha}} (E_{\nu}),
\end{equation}
where $d$ is the distance to the source; $\lambda_{\gamma \gamma}$ is the interaction length of gamma rays with radiation backgrounds; $K=1$ for photo-hadronic interactions and $K=2$ for hadro-nuclear interactions; the sum is over the neutrino flavors. Using the neutrino flux normalization, and assuming photo-hadronic interactions, we can obtain the gamma-ray flux normalization from Eq. \ref{eq:nuem}\footnote{For our estimation, and to compare with the result from \cite{amonHAWCIC}, we assume an equal ratio of neutrino flavors when detected. This makes all three flavor fluxes similar and hence the factor of 1/3 is canceled. Since $E_{\gamma}\approx2E_{\nu}$ we end up with $F_{\gamma}(E_{\gamma}) \approx F_{\nu}(E_{\nu})$ at location $d=0$. The gamma-ray flux is then attenuated as mentioned in the main text.}.

After obtaining the gamma-ray flux normalization, we inject the sources in HAWC simulated data. Using the simulated redshift information, we apply the attenuation of gamma rays from the extragalactic background light using the model from \cite{eblDominguez}. After running the HAWC analysis, if the observed hotspot has a significance larger than 2.75$\sigma$, we proceed to inject the neutrinos using Monte Carlo signal data from the ANTARES simulation as well as background events from the ANTARES scrambled data sets. Then we proceed to calculate the RS as explained in \S\ref{sec:statistic}. We simulate transient sources for a period with the same livetime as the archival data being used in this analysis. 

To be able to estimate the sensitivity and discovery potential, we use the number of coincidences above the 1 per year threshold as a statistic. For the livetime of the analysis, we expect to observe at least $\sim 4$ random coincidences. Using random samples from the RS distribution, we find that the distribution of the number of coincidences behaves as a Poisson distribution with a mean of $\lambda_{_{\rm bk}} = 4.39$, since that is the livetime of the analysis.
We now need to find limits on the total signal and background rate , $\lambda_{_{\rm bk}} + \lambda_{_{\rm s}}$.
In order to obtain the sensitivity we need a total rate that will produce a Poisson distribution with 90\% of its population above the median of the background Poisson distribution. For the discovery potential, we need a total rate that will produce a Poisson distribution with 50\% of its population above the threshold of the $p$-value $=2.87\times 10^{-7}$ (5$\sigma$) of the background distribution. The mean signal $\lambda_{_{\rm s}}$ values for the sensitivity and discovery potential are 3.6 and 14.3. 

We perform the simulation 100 times for each pair of local rate density of transient sources and total isotropic equivalent energies to gather enough statistics to build the Poisson distributions. The value of the rate densities and isotropic energies that give the desired $\lambda_{_{\rm s}}$ values are shown in Figure \ref{fig:sensitivity}. The figure also shows several populations of transient sources, with a range of local rate densities and isotropic energies obtained from \cite{mmsources}. We see that long gamma-ray bursts are the only sources from which we may expect some detectable coincidences.

\section{Archival Results}
\begin{figure}
    \centering
    \includegraphics[scale=0.55]{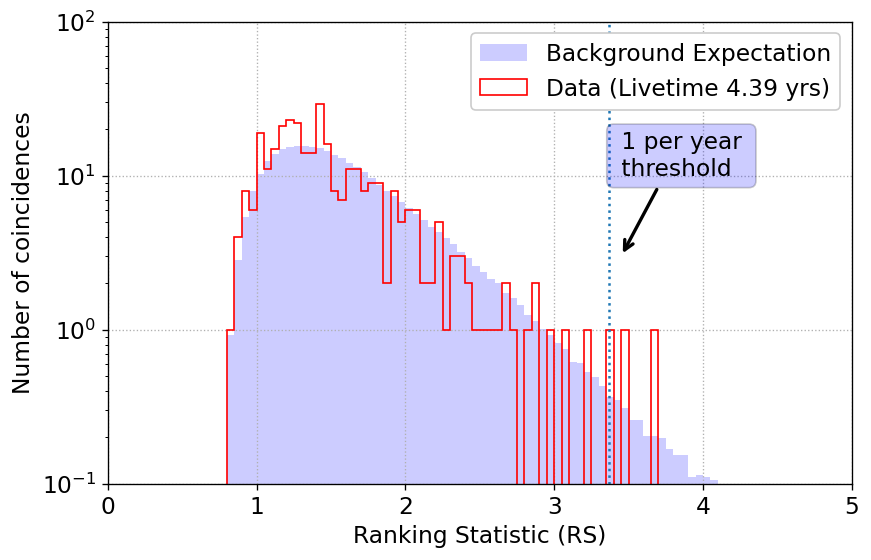}
    \caption{Ranking statistic distribution of the analysis. Blue: expected distribution obtained from the scrambled data set and normalized to the number of coincidences observed in the data set. Red: distribution of the unscrambled data. The vertical line marks the 1-per-year FAR coincidence threshold.}
    \label{fig:rankDist}
\end{figure}

\begin{figure}
    \centering
    \includegraphics[scale=0.5]{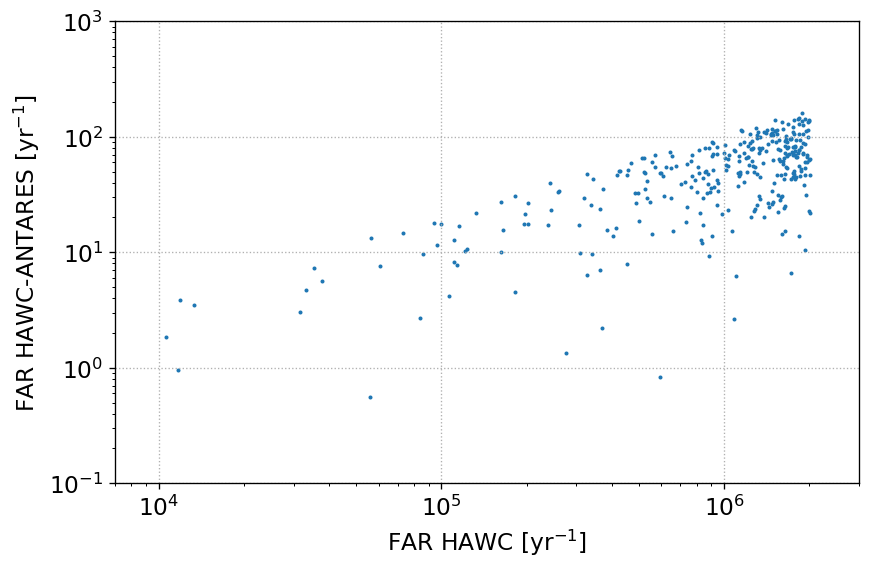}
    \caption{Comparison of the FAR of the coincidence analysis  vs the FAR of HAWC alone. The combined analysis reduces by several orders of magnitude the FAR of the events.}
    \label{fig:farComp}
\end{figure}

The ranking distribution for the archival data is shown in Figure \ref{fig:rankDist}, along with the simulated distribution from the scrambled data set used in Section \ref{sec:far} to obtain the FAR. 
The power of the combined data analysis can be seen in Figure \ref{fig:farComp}, which shows the FAR for the HAWC-ANTARES coincidences versus the FAR for the HAWC events alone. As expected, the FAR for HAWC only events is reduced by 4 orders of magnitude. The low FAR of coincidences makes it useful for follow-up searches in real time.

After performing the analysis on unscrambled data, we found 3 events that pass the 1 per year FAR threshold in this period. These three events are summarized in Table \ref{tab:Coincs}. Although these events are rare given their FAR, they are still consistent with background as shown by the post-trials $p$-value (last column in Table \ref{tab:Coincs}, calculated as $p_{\rm post-trials} = 1-\exp({-{\rm FAR}\cdot \Delta T})$, where $\Delta T$ is the livetime of the analysis).  Tables \ref{tab:CoincHAWC} and \ref{tab:CoincANT} have the information of the events that make the coincidences. 

\begin{deluxetable*}{ccccccc}
\tablecaption{Summary information for the three coincidences with FAR $<1 {\rm~yr}^{-1}$. Information for the HAWC and ANTARES events that make these coincidences are found in Tables \ref{tab:CoincHAWC} and \ref{tab:CoincANT}. The coincidence positions and uncertainty circles are shown as red in the skymaps of Figure \ref{fig:skymaps}. \label{tab:Coincs}}
\tabletypesize{\scriptsize}
\tablehead{
\colhead{Coincidence ID} & \colhead{Dec [deg]} & \colhead{RA [deg]} & \colhead{$U_C$ (50\%)[deg]} & \colhead{RS} & \colhead{FAR [${\rm yr}^{-1}$]} & \colhead{$p$-value} 
}
\startdata
1 & 25.0 & 25.6 & 0.18 &  3.46 & 0.83 & 0.97 \\
\hline
2 & -0.8 & 222.7 & 0.16 & 3.38 & 0.96 & 0.98\\
\hline
3 & 3.4 & 85.4 & 0.16 & 3.65 & 0.56 & 0.91\\
\enddata
\tablenotetext{}{\textbf{Note. }The uncertainty $U_C$(50\%) corresponds to the 50\% containment region of the estimated position of the coincidence. RS is the ranking statistic as defined in Section \ref{sec:statistic}. The $p$-value corresponds to the post-trial $p$-value.}
\end{deluxetable*}

\begin{deluxetable*}{cccccccc}
\tablecaption{Information on the HAWC ``hotspots'' that correspond to each of the coincidences with a FAR$<1 {\rm ~yr}^{-1}$ per year. The positions and uncertainty circles of the HAWC ``hotspots'' are shown as blue in the skymaps of Figure \ref{fig:skymaps}\label{tab:CoincHAWC}. }
\tabletypesize{\scriptsize}
\tablehead{
\colhead{Dec} & \colhead{RA} & \colhead{$U_H$(50\%)} & \colhead{Rising Time} & \colhead{Setting Time} & \colhead{Significance} & \colhead{Flux} & \colhead{Spectral Index} \\
\colhead{[deg]} & \colhead{[deg]} & \colhead{[deg]} & \colhead{[UT]} & \colhead{[UT]} & $\sigma$ & \colhead{$\times 10^{-11}$[TeV$^{-1}$ cm$^{-2}$ s$^{-1}$]} & \colhead{}
}
\startdata
25.2 & 25.7 & 0.20 & 2016-01-07 21:29:40 & 2016-01-08 04:39:38 & 3.18 & 2.0$\pm$0.8 & 2.5 \\
\hline
\hline
-0.8 & 222.4 & 0.20 & 2017-09-06 19:08:16 & 2017-09-07 01:21:22 & 4.29 & 5.0$\pm$1.6 & 2.5  \\
\hline
\hline
3.4 & 85.7 & 0.17 & 2019-03-28 20:33:04 & 2019-03-29 03:01:18 & 3.89 & 4.9$\pm$1.6 & 3.0  \\
\enddata
\tablenotetext{}{\textbf{Note.} The uncertainty $U_H$(50\%) corresponds to the 50\% containment region of the HAWC hotspot. The assumed flux model is a power law with an index shown in the last column. The index is fixed during the fit. The flux measurement is the normalization of the power law at 1 TeV.}
\end{deluxetable*}

\begin{deluxetable*}{cccccc}
\tablecaption{ANTARES event information for each coincidence. The positions and uncertainty circles of the ANTARES events are shown as black in the skymaps of Figure \ref{fig:skymaps}\label{tab:CoincANT}.}
\tabletypesize{\scriptsize}
\tablehead{
\colhead{Dec} & \colhead{RA} & \colhead{$U_A$(50\%)} & \colhead{Time} & \colhead{Background $p$-value} & \colhead{$\Delta \theta$} \\
\colhead{[deg]} & \colhead{[deg]} & \colhead{[deg]} & \colhead{[UT]} & \colhead{}& \colhead{[deg]} 
}
\startdata
24.1 & 25.4 & 0.45 & 2016-01-08 04:24:40.32 & 0.009 & 1.10\\
\hline
\hline
-0.5 & 225.6 & 0.47 & 2017-09-06 22:10:24.96 & 0.095 & 3.28\\
\hline
\hline
3.4 & 85.6 & 0.36 & 2019-03-29 01:03:47.0 & 0.51 & 0.33 \\
\enddata
\tablenotetext{}{\textbf{Note.} The uncertainty $U_A$(50\%) corresponds to the 50\% containment region of the ANTARES position. $\Delta \theta$ is the distance from the best-fit HAWC hotspot position to the measured ANTARES event position.}
\end{deluxetable*}

\begin{figure}%
	\centering
	\gridline{\fig{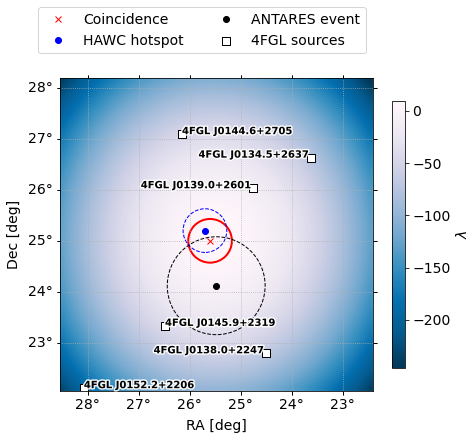}{0.4\textwidth}{(a) Coincidence 1.}
	          \fig{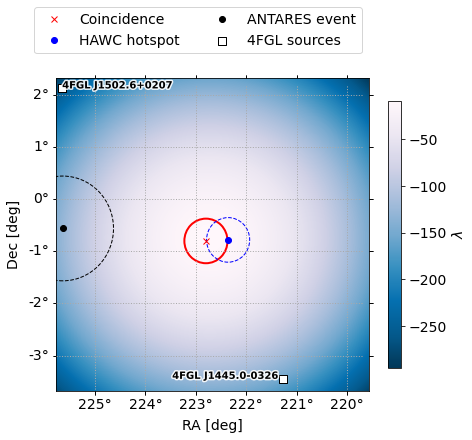}{0.4\textwidth}{(b) Coincidence 2.}}
	\gridline{\fig{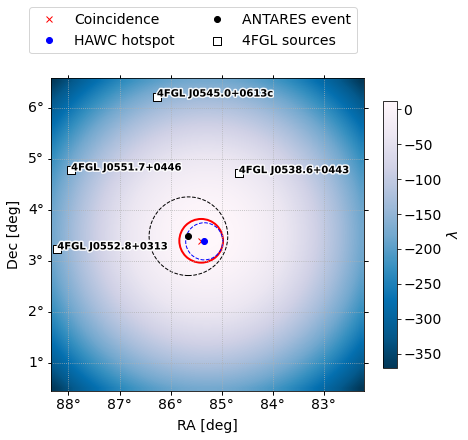}{0.4\textwidth}{(c) Coincidence 3.}}
	\caption{Sky maps in celestial coordinates of the HAWC-ANTARES coincidences with FAR values below 1 coincidence per year found in the archival data. The positions of the individual events are marked with the dots. The best-fit combined positions $\boldsymbol{x}_\mathrm{coinc}$,  found after optimizing $\lambda( \bm{x} )$ (Eq. \ref{eq:lambda}), are marked with a cross. Circles represent the 50\% containment regions.}
	\label{fig:skymaps}
\end{figure}

The sky maps of the coincidences are shown in Figure \ref{fig:skymaps}. Each sky map shows the position of the individual events along with their uncertainties, as well as the best position of the coincidence. Also shown are the sources of the 4FGL Catalog \citep{4fgl} that appear in each of the regions.

We searched for past activity around the coincidences by looking in the \textit{Fermi} All-sky Variability Analysis (FAVA) online tool\footnote{\url{https://fermi.gsfc.nasa.gov/ssc/data/access/lat/FAVA/}}. We did not find any past activity in the regions of the coincidences 2 and 3. For coincidence 1,  the source J0144.6+2705, associated with TXS 0141+268, is located 2.0$^{\circ}$ away from the position of the coincidence. FAVA reported a burst in 2018 from this source, which was found in the high-energy band (800 MeV$-$300 GeV).

A search in the SIMBAD Catalog \citep{simbad} revealed several sources inside the uncertainty regions of the positions of the coincidences. For coincidence 1, we found several quasars, along with a radio source and an X-ray source. All of the quasars have redshift measurements larger than 0.3, the farthest HAWC can observe before the gamma rays start to be severely attenuated by the extragalactic background light \citep{hawc_agn}. 

In coincidence 2, there were 115 sources inside the uncertainty region of the coincidence in the SIMBAD Catalog. Around 15 sources are stars, while the rest are galaxies. 

For coincidence 3 in 2019, we found only 8 sources in the SIMBAD catalog: 3 molecular clouds, 2 stars, 2 radio sources and one X-ray source. No information about the distance for the radio sources or X-ray source were available. 

These coincidences are examples where, if the analysis had been running in real time, a follow-up observation in another wavelength could have pinpointed any source that is flaring in the region.

\subsection{Upper Limit}
After observing 3 coincidences in 4.39 years of data with a FAR of less than 1 per year, we calculate the 90\% confidence level upper limit for the parameter space presented in Figure \ref{fig:sensitivity}. Using Equation (9.54) from \cite{analysisbook}, we obtain $\lambda_{_{\rm signal}}=3.85$. We apply the procedure of Section \ref{sec:sensitivity} to find the upper limit on the total isotropic equivalent neutrino energy as a function of the local source rate. 


\section{Conclusion}

Archival data that span between 2015 and 2020 was analyzed to search for multimessenger sources through a coincidence analysis between subthreshold data of the HAWC observatory and public data from the ANTARES neutrino telescope. In this time period, three coincidences were found with a FAR of less than one coincidence per year. Although these coincidences are consistent with background expectations, they are still useful for follow-up observations, since the FAR can be improved by several orders of magnitude, compared to when the events are coming from the individual detectors. It is possible that a flare in the region could have been observed by a follow-up telescope, hinting at the presence of a multimessenger source, if the analysis had been running in real time.

Furthermore, based on the sensitivity and discovery potential studies, we found that long gamma-ray bursts are potential candidates for a possible coincidence detection with this analysis. 
This work is also a proof of principle analysis for future neutrino observatories. In this sense, with the end of operations of ANTARES, we expect that this analysis will be implemented for KM3NeT with an already exceeding ANTARES effective volume. Based on the information in \cite{km3net}, a back-of-the-envelope estimate suggests an improvement of the sensitivity and discovery potential of more than one order of magnitude, assuming the same livetime. We look forward to implementing this new stream within AMON.

\begin{acknowledgments}

AMON: This research or portions of this research were conducted with Advanced CyberInfrastructure computational resources provided by the Institute for Computational and Data Sciences at the Pennsylvania State University ( https://www.icds.psu.edu/ ).
This material is based upon work supported by the National Science Foundation under Grants PHY-1708146 and PHY-1806854 and by the Institute for Gravitation and the Cosmos of the Pennsylvania State University. Any opinions, findings, and conclusions or recommendations expressed in this material are those of the author(s) and do not necessarily reflect the views of the National Science Foundation. 

ANTARES: The authors acknowledge the financial support of the funding agencies:
Centre National de la Recherche Scientifique (CNRS), Commissariat \`a
l'\'ener\-gie atomique et aux \'energies alternatives (CEA),
Commission Europ\'eenne (FEDER fund and Marie Curie Program),
Institut Universitaire de France (IUF), LabEx UnivEarthS (ANR-10-LABX-0023 and ANR-18-IDEX-0001),
R\'egion \^Ile-de-France (DIM-ACAV), R\'egion
Alsace (contrat CPER), R\'egion Provence-Alpes-C\^ote d'Azur,
D\'e\-par\-tement du Var and Ville de La
Seyne-sur-Mer, France;
Bundesministerium f\"ur Bildung und Forschung
(BMBF), Germany; 
Istituto Nazionale di Fisica Nucleare (INFN), Italy;
Nederlandse organisatie voor Wetenschappelijk Onderzoek (NWO), the Netherlands;
Executive Unit for Financing Higher Education, Research, Development and Innovation (UEFISCDI), Romania;
Ministerio de Ciencia, Innovaci\'{o}n, Investigaci\'{o}n y
Universidades (MCIU): Programa Estatal de Generaci\'{o}n de
Conocimiento (refs. PGC2018-096663-B-C41, -A-C42, -B-C43, -B-C44
and refs. PID2021-124591NB-C41, -C42, -C43)
(MCIU/FEDER), Generalitat Valenciana: Prometeo (PROMETEO/2020/019),
Grisol\'{i}a (refs. GRISOLIA/2018/119, /2021/192) and GenT
(refs. /2019/043, /2020/049, /2021/023) programs, Junta de
Andaluc\'{i}a (ref. A-FQM-053-UGR18), La Caixa Foundation (ref. LCF/BQ/IN17/11620019), EU: MSC program (ref. 101025085), Spain;
Ministry of Higher Education, Scientific Research and Innovation, Morocco, and the Arab Fund for Economic and Social Development, Kuwait.
We also acknowledge the technical support of Ifremer, AIM and Foselev Marine
for the sea operation and the CC-IN2P3 for the computing facilities.

HAWC: We acknowledge the support from: the US National Science Foundation (NSF); the US Department of Energy Office of High-Energy Physics; the Laboratory Directed Research and Development (LDRD) program of Los Alamos National Laboratory; Consejo Nacional de Ciencia y Tecnolog\'ia (CONACyT), M\'exico, grants 271051, 232656, 260378, 179588, 254964, 258865, 243290, 132197, A1-S-46288, A1-S-22784, c\'atedras 873, 1563, 341, 323, Red HAWC, M\'exico; DGAPA-UNAM grants IG101320, IN111716-3, IN111419, IA102019, IN110621, IN110521; VIEP-BUAP; PIFI 2012, 2013, PROFOCIE 2014, 2015; the University of Wisconsin Alumni Research Foundation; the Institute of Geophysics, Planetary Physics, and Signatures at Los Alamos National Laboratory; Polish Science Centre grant, DEC-2017/27/B/ST9/02272; Coordinaci\'on de la Investigaci\'on Cient\'ifica de la Universidad Michoacana; Royal Society - Newton Advanced Fellowship 180385; Generalitat Valenciana, grant CIDEGENT/2018/034; The Program Management Unit for Human Resources \& Institutional Development, Research and Innovation, NXPO (grant number B16F630069); Coordinaci\'on General Acad\'emica e Innovaci\'on (CGAI-UdeG), PRODEP-SEP UDG-CA-499; Institute of Cosmic Ray Research (ICRR), University of Tokyo, H.F. acknowledges support by NASA under award number 80GSFC21M0002. We also acknowledge the significant contributions over many years of Stefan Westerhoff, Gaurang Yodh and Arnulfo Zepeda Dominguez, all deceased members of the HAWC collaboration. Thanks to Scott Delay, Luciano D\'iaz and Eduardo Murrieta for technical support.

\end{acknowledgments}

\facilities{HAWC, ANTARES, AMON}

\software{astropy~\citep{astropy},
        FIRESONG~\citep{firesong},
        numpy~\citep{numpy},
        scipy~\citep{scipy}
        matplotlib~\citep{matplotlib},
        pandas~\citep{pandas},
        amonpy~\citep{amon2020}
        }

\bibliography{biblio}{}
\bibliographystyle{aasjournal}



\end{document}

%% file: authors.tex
\author[0000-0002-2084-5049]{H.A.~Ayala Solares}
\affiliation{Department of Physics, Pennsylvania State University, University Park, PA 16802, USA}

\author[0000-0003-2923-2246]{S.~Coutu}
\affiliation{Department of Physics, Pennsylvania State University, University Park, PA 16802, USA}

\author[0000-0003-4738-0787]{D. Cowen}
\affiliation{Department of Physics, Pennsylvania State University, University Park, PA 16802, USA}


\author[0000-0002-3714-672X]{D.~B. Fox}
\affiliation{Department of Physics, Pennsylvania State University, University Park, PA 16802, USA}
\author[0000-0001-8711-1456]{T. Gr{\'e}goire}
\affiliation{Department of Physics, Pennsylvania State University, University Park, PA 16802, USA}


\author[0000-0001-6191-1244]{F.~McBride}
\affiliation{Department of Physics and Astronomy, Bowdoin College, Brunswick, Maine 04011, USA}

\author[0000-0002-7675-4656]{M.~Mostaf\'{a}}
\affiliation{Department of Physics, Pennsylvania State University, University Park, PA 16802, USA}

\author[0000-0002-5358-5642]{K.~Murase}
\affiliation{Department of Physics, Pennsylvania State University, University Park, PA 16802, USA}

\author[0000-0003-0569-6978]{S.~Wissel}
\affiliation{Department of Physics, Pennsylvania State University, University Park, PA 16802, USA}

\collaboration{12}{AMON Team}%

\author{A.~Albert}
\affiliation{\scriptsize{Universit\'e de Strasbourg, CNRS,  IPHC UMR 7178, F-67000 Strasbourg, France}}
\affiliation{\scriptsize Universit\'e de Haute Alsace, F-68100 Mulhouse, France}

\author{S.~Alves}
\affiliation{\scriptsize{IFIC - Instituto de F\'isica Corpuscular (CSIC - Universitat de Val\`encia) c/ Catedr\'atico Jos\'e Beltr\'an, 2 E-46980 Paterna, Valencia, Spain}}

\author{M.~Andr\'e}
\affiliation{{\scriptsize{Technical University of Catalonia, Laboratory of Applied Bioacoustics, Rambla Exposici\'o, 08800 Vilanova i la Geltr\'u, Barcelona, Spain}}}

\author[0000-0002-3199-594X]{M.~Ardid}
\affiliation{\scriptsize{Institut d'Investigaci\'o per a la Gesti\'o Integrada de les Zones Costaneres (IGIC) - Universitat Polit\`ecnica de Val\`encia. C/  Paranimf 1, 46730 Gandia, Spain}}

\author[0000-0003-4821-6655]{S.~Ardid}
\affiliation{\scriptsize{Institut d'Investigaci\'o per a la Gesti\'o Integrada de les Zones Costaneres (IGIC) - Universitat Polit\`ecnica de Val\`encia. C/  Paranimf 1, 46730 Gandia, Spain}}

\author{J.-J.~Aubert}
\affiliation{\scriptsize{Aix Marseille Univ, CNRS/IN2P3, CPPM, Marseille, France}}

\author{J.~Aublin}
\affiliation{\scriptsize{Universit\'e de Paris, CNRS, Astroparticule et Cosmologie, F-75013 Paris, France}}

\author[0000-0001-6064-3858]{B.~Baret}
\affiliation{\scriptsize{Universit\'e de Paris, CNRS, Astroparticule et Cosmologie, F-75013 Paris, France}}

\author{S.~Basa}
\affiliation{\scriptsize{Aix Marseille Univ, CNRS, CNES, LAM, Marseille, France }}

\author[0000-0001-6064-3858]{B.~Belhorma}
\affiliation{\scriptsize{National Center for Energy Sciences and Nuclear Techniques, B.P.1382, R. P.10001 Rabat, Morocco}}

% ...

\collaboration{200}{ANTARES Collaboration}

\author[0000-0003-0197-5646]{A.~Albert}
\affiliation{Physics Division, Los Alamos National Laboratory, Los Alamos, NM, USA }


\author{C.~Alvarez}
\affiliation{Universidad Aut\'{o}noma de Chiapas, Tuxtla Guti\'{e}rrez, Chiapas, M\'{e}xico}


\author{J.C.~Arteaga-Vel\'{a}zquez}
\affiliation{Universidad Michoacana de San Nicol\'{a}s de Hidalgo, Morelia, M\'{e}xico }




\author[0000-0002-5529-6780]{R.~Babu}
\affiliation{Department of Physics, Michigan Technological University, Houghton, MI, USA}

\author[0000-0003-3207-105X]{E.~Belmont-Moreno}
\affiliation{Instituto de F\'{i}sica, Universidad Nacional Aut\'{o}noma de M\'{e}xico, Ciudad de M\'{e}xico, M\'{e}xico }


\author[0000-0002-4042-3855]{K.S.~Caballero-Mora}
\affiliation{Universidad Aut\'{o}noma de Chiapas, Tuxtla Guti\'{e}rrez, Chiapas, M\'{e}xico}

\author[0000-0003-2158-2292]{T.~Capistr\'n}
\affiliation{Instituto de Astronom\'{i}a, Universidad Nacional Aut\'{o}noma de M\'{e}xico, Ciudad de M\'{e}xico, M\'{e}xico }

\author[0000-0002-8553-3302]{A.~Carrami\~{n}ana}
\affiliation{Instituto Nacional de Astrof\'{i}sica, \'{O}ptica y Electr\'{o}nica, Puebla, M\'{e}xico }

\author[0000-0002-6144-9122]{S.~Casanova}
\affiliation{Institute of Nuclear Physics Polish Academy of Sciences, PL-31342 IFJ-PAN, Krakow, Poland }

\author[0000-0002-7607-9582]{U.~Cotti}
\affiliation{Universidad Michoacana de San Nicol\'{a}s de Hidalgo, Morelia, M\'{e}xico }

\author{O.~Chaparro-Amaro}
\affiliation{Centro de Investigaci\'on en Computaci\'on, Instituto Polit\'ecnico Nacional, Mexico City, Mexico }

\author[0000-0002-1132-871X]{J.~Cotzomi}
\affiliation{Facultad de Ciencias F\'{i}sico Matemáticas, Benem\'{e}rita Universidad Aut\'{o}noma de Puebla, Puebla, M\'{e}xico }

\author[0000-0002-7747-754X]{S.~Couti\~no de Le\'{o}n}
\affiliation{Department of Physics, University of Wisconsin-Madison, Madison, WI 53706, USA }

\author[0000-0001-9643-4134]{E.~De la Fuente}
\affiliation{Departamento de F\'{i}sica, Centro Universitario de Ciencias Exactas e Ingenierias, Universidad de Guadalajara, Guadalajara, M\'{e}xico }

\author[0000-0002-8528-9573]{C.~de Le\'{o}n}
\affiliation{Universidad Michoacana de San Nicol\'{a}s de Hidalgo, Morelia, M\'{e}xico }

\author{R.~Diaz Hernandez}
\affiliation{Instituto Nacional de Astrof\'{i}sica, \'{O}ptica y Electr\'{o}nica, Puebla, M\'{e}xico }


\author[0000-0002-2987-9691]{M.A.~DuVernois}
\affiliation{Department of Physics, University of Wisconsin-Madison, Madison, WI 53706, USA }

\author[0000-0003-2169-0306]{M.~Durocher}
\affiliation{Physics Division, Los Alamos National Laboratory, Los Alamos, NM, USA }

\author[0000-0002-0087-0693]{J.C.~D\'{i}az-V\'{e}lez}
\affiliation{Departamento de F\'{i}sica, Centro Universitario de Ciencias Exactas e Ingenierias, Universidad de Guadalajara, Guadalajara, M\'{e}xico }

\author{K.~Engel}
\affiliation{Dept. of Physics, University of Maryland, College Park, MD 20742, USA}

\author[0000-0001-7074-1726]{C.~Espinoza}
\affiliation{Instituto de F\'{i}sica, Universidad Nacional Aut\'{o}noma de M\'{e}xico, Ciudad de M\'{e}xico, M\'{e}xico }

\author[0000-0002-8246-4751]{K.L.~Fan}
\affiliation{Dept. of Physics, University of Maryland, College Park, MD 20742, USA}

\author{M.~Fern\'{a}ndez Alonso}
\affiliation{Department of Physics, Pennsylvania State University, University Park, PA 16802, USA}


\author[0000-0002-0173-6453]{N.~Fraija}
\affiliation{Instituto de Astronom\'{i}a, Universidad Nacional Aut\'{o}noma de M\'{e}xico, Ciudad de M\'{e}xico, M\'{e}xico }



\author[0000-0002-4188-5584]{J.A.~Garc\'{i}a-Gonz\'{a}lez}
\affiliation{Instituto de Astronom\'{i}a, Universidad Nacional Aut\'{o}noma de M\'{e}xico, Ciudad de M\'{e}xico, M\'{e}xico }

\author[0000-0003-1122-4168]{F.~Garfias}
\affiliation{Instituto de Astronom\'{i}a, Universidad Nacional Aut\'{o}noma de M\'{e}xico, Ciudad de M\'{e}xico, M\'{e}xico }

\author[0000-0002-5209-5641]{M.M.~Gonz\'{a}lez}
\affiliation{Instituto de Astronom\'{i}a, Universidad Nacional Aut\'{o}noma de M\'{e}xico, Ciudad de M\'{e}xico, M\'{e}xico }

\author[0000-0002-9790-1299]{J.A.~Goodman}
\affiliation{Dept. of Physics, University of Maryland, College Park, MD 20742, USA}

\author[0000-0001-9844-2648]{J.P.~Harding}
\affiliation{Physics Division, Los Alamos National Laboratory, Los Alamos, NM, USA }

\author{S.~Hernandez}
\affiliation{Instituto de F\'{i}sica, Universidad Nacional Aut\'{o}noma de M\'{e}xico, Ciudad de M\'{e}xico, M\'{e}xico }


\author[0000-0002-3808-4639]{D.~Huang}
\affiliation{Department of Physics, Michigan Technological University, Houghton, MI, USA }

\author[0000-0002-5527-7141]{F.~Hueyotl-Zahuantitla}
\affiliation{Universidad Aut\'{o}noma de Chiapas, Tuxtla Guti\'{e}rrez, Chiapas, M\'{e}xico}

\author{P.~H{\"u}ntemeyer}
\affiliation{Department of Physics, Michigan Technological University, Houghton, MI, USA }

\author[0000-0001-5811-5167]{A.~Iriarte}
\affiliation{Instituto de Astronom\'{i}a, Universidad Nacional Aut\'{o}noma de M\'{e}xico, Ciudad de M\'{e}xico, M\'{e}xico }


\author[0000-0003-4467-3621]{V.~Joshi}
\affiliation{Erlangen Centre for Astroparticle Physics, Friedrich-Alexander-Universit{\"a}t Erlangen-N{\"u}rnberg, D-91058 Erlangen, Germany}

\author{S.~Kaufmann}
\affiliation{Universidad Politecnica de Pachuca, Pachuca, Hgo, M\'{e}xico }

\author{A.~Lara}
\affiliation{Instituto de Geof\'{i}sica, Universidad Nacional Aut\'{o}noma de M\'{e}xico, Ciudad de M\'{e}xico, M\'{e}xico }

\author[0000-0001-5516-4975]{H.~Le\'{o}n Vargas}
\affiliation{Instituto de F\'{i}sica, Universidad Nacional Aut\'{o}noma de M\'{e}xico, Ciudad de M\'{e}xico, M\'{e}xico }

\author[0000-0003-2696-947X]{J.T.~Linnemann}
\affiliation{Dept. of Physics and Astronomy, Michigan State University, East Lansing, MI 48824, USA}

\author[0000-0001-8825-3624]{A.L.~Longinotti}
\affiliation{Instituto Nacional de Astrof\'{i}sica, \'{O}ptica y Electr\'{o}nica, Puebla, M\'{e}xico }

\author[0000-0003-2810-4867]{G.~Luis-Raya}
\affiliation{Universidad Politecnica de Pachuca, Pachuca, Hgo, M\'{e}xico }


\author[0000-0001-8088-400X]{K.~Malone}
\affiliation{Space Science and Applications Group, Los Alamos National Laboratory, Los Alamos, NM, USA }

\author[0000-0001-9052-856X]{O.~Martinez}
\affiliation{Facultad de Ciencias F\'{i}sico Matemáticas, Benem\'{e}rita Universidad Aut\'{o}noma de Puebla, Puebla, M\'{e}xico }

\author[0000-0001-9035-1290]{I.~Martinez-Castellanos}
\affiliation{Dept. of Physics, University of Maryland, College Park, MD 20742, USA}

\author[0000-0002-2824-3544]{J.~Mart\'{i}nez-Castro}
\affiliation{Centro de Investigaci\'on en Computaci\'on, Instituto Polit\'ecnico Nacional, Mexico City, Mexico.}

\author[0000-0002-2610-863X]{J.A.~Matthews}
\affiliation{Dept of Physics and Astronomy, University of New Mexico, Albuquerque, NM, USA }

\author[0000-0002-8390-9011]{P.~Miranda-Romagnoli}
\affiliation{Universidad Aut\'{o}noma del Estado de Hidalgo, Pachuca, M\'{e}xico }

\author{J.A.~Morales-Soto}
\affiliation{Universidad Michoacana de San Nicol\'{a}s de Hidalgo, Morelia, M\'{e}xico }

\author[0000-0002-1114-2640]{E.~Moreno}
\affiliation{Facultad de Ciencias F\'{i}sico Matemáticas, Benem\'{e}rita Universidad Aut\'{o}noma de Puebla, Puebla, M\'{e}xico }


\author{A.~Nayerhoda}
\affiliation{Institute of Nuclear Physics Polish Academy of Sciences, PL-31342 IFJ-PAN, Krakow, Poland }

\author[0000-0003-1059-8731]{L.~Nellen}
\affiliation{Instituto de Ciencias Nucleares, Universidad Nacional Aut\'{o}noma de M\'{e}xico, Ciudad de M\'{e}xico, M\'{e}xico }


\author[0000-0002-6859-3944]{M.U.~Nisa}
\affiliation{Dept. of Physics and Astronomy, Michigan State University, East Lansing, MI 48824, USA}

\author[0000-0001-7099-108X]{R.~Noriega-Papaqui}
\affiliation{Universidad Aut\'{o}noma del Estado de Hidalgo, Pachuca, M\'{e}xico }

\author{N.~Omodei}
\affiliation{Department of Physics, Stanford University, Stanford, CA 94305-4060, USA}

\author{A.~Peisker}
\affiliation{Dept. of Physics and Astronomy, Michigan State University, East Lansing, MI 48824, USA}

\author{Y.~P\'{e}rez Araujo}
\affiliation{Instituto de Astronom\'{i}a, Universidad Nacional Aut\'{o}noma de M\'{e}xico, Ciudad de M\'{e}xico, M\'{e}xico }

\author[0000-0001-5998-4938]{E.G.~P\'{e}rez-P\'{e}rez}
\affiliation{Universidad Politecnica de Pachuca, Pachuca, Hgo, M\'{e}xico }

\author[0000-0002-6524-9769]{C.D.~Rho}
\affiliation{Natural Science Research Institute, University of Seoul, Seoul, Republic of Korea}

\author[0000-0003-1327-0838]{D.~Rosa-Gonz\'{a}lez}
\affiliation{Instituto Nacional de Astrof\'{i}sica, \'{O}ptica y Electr\'{o}nica, Puebla, M\'{e}xico }

\author{E.~Ruiz-Velasco}
\affiliation{Max-Planck Institute for Nuclear Physics, 69117 Heidelberg, Germany}

\author{H.~Salazar}
\affiliation{Facultad de Ciencias F\'{i}sico Matem\'{a}ticas, Benem\'{e}rita Universidad Aut\'{o}noma de Puebla, Puebla, M\'{e}xico }

\author[0000-0002-8610-8703]{F.~Salesa Greus}
\affiliation{Institute of Nuclear Physics Polish Academy of Sciences, PL-31342 IFJ-PAN, Krakow, Poland }
\affiliation{\scriptsize{IFIC - Instituto de F\'isica Corpuscular (CSIC - Universitat de Val\`encia) c/ Catedr\'atico Jos\'e Beltr\'an, 2 E-46980 Paterna, Valencia, Spain}}

\author[0000-0001-6079-2722]{A.~Sandoval}
\affiliation{Instituto de F\'{i}sica, Universidad Nacional Aut\'{o}noma de M\'{e}xico, Ciudad de M\'{e}xico, M\'{e}xico }

\author[0000-0001-8644-4734]{M.~Schneider}
\affiliation{Dept. of Physics, University of Maryland, College Park, MD 20742, USA}

\author[0000-0002-1012-0431]{A.J.~Smith}
\affiliation{Dept. of Physics, University of Maryland, College Park, MD 20742, USA}

\author{Y.~Son}
\affiliation{Natural Science Research Institute, University of Seoul, Seoul, Republic of Korea}

\author[0000-0002-1492-0380]{R.W.~Springer}
\affiliation{Department of Physics and Astronomy, University of Utah, Salt Lake City, UT, USA }

\author{O.~Tibolla}
\affiliation{Universidad Politecnica de Pachuca, Pachuca, Mexico}

\author[0000-0001-9725-1479]{K.~Tollefson}
\affiliation{Dept. of Physics and Astronomy, Michigan State University, East Lansing, MI 48824, USA}

\author[0000-0002-1689-3945]{I.~Torres}
\affiliation{Instituto Nacional de Astrof\'{i}sica, \'{O}ptica y Electr\'{o}nica, Puebla, M\'{e}xico }

\author{R.~Torres-Escobedo}
\affiliation{Departamento de F\'{i}sica, Centro Universitario de Ciencias Exactas e Ingenierias, Universidad de Guadalajara, Guadalajara, M\'{e}xico }

\author{R.~Turner}
\affiliation{Department of Physics, Michigan Technological University, Houghton, MI, USA }

\author[0000-0002-2748-2527]{F.~Ure\~{n}a-Mena}
\affiliation{Instituto Nacional de Astrof\'{i}sica, \'{O}ptica y Electr\'{o}nica, Puebla, M\'{e}xico }

\author{E.~Varela}
\affiliation{Facultad de Ciencias F\'{i}sico Matemáticas, Benem\'{e}rita Universidad Aut\'{o}noma de Puebla, Puebla, M\'{e}xico }


\author{X.~Wang}
\affiliation{Department of Physics, Michigan Technological University, Houghton, MI, USA }


\author[0000-0003-0625-6675]{K.~Whitaker}
\affiliation{Department of Physics, Pennsylvania State University, University Park, PA 16802, USA}

\author[0000-0002-6623-0277]{E.~Willox}
\affiliation{Dept. of Physics, University of Maryland, College Park, MD 20742, USA}

\author[0000-0001-9976-2387]{A.~Zepeda}
\affiliation{Physics Department, Centro de Investigacion y de Estudios Avanzados del IPN, Mexico City, DF, Mexico }

\author[0000-0003-0513-3841]{H.~Zhou}
\affiliation{Tsung-Dao Lee Institute \& School of Physics and Astronomy, Shanghai Jiao Tong University, Shanghai, China}

\collaboration{200}{HAWC Collaboration}

%% file: main.bbl
\begin{thebibliography}{}
\expandafter\ifx\csname natexlab\endcsname\relax\def\natexlab#1{#1}\fi
\providecommand{\url}[1]{\href{#1}{#1}}
\providecommand{\dodoi}[1]{doi:~\href{http://doi.org/#1}{\nolinkurl{#1}}}
\providecommand{\doeprint}[1]{\href{http://ascl.net/#1}{\nolinkurl{http://ascl.net/#1}}}
\providecommand{\doarXiv}[1]{\href{https://arxiv.org/abs/#1}{\nolinkurl{https://arxiv.org/abs/#1}}}

\bibitem[{Abdollahi {et~al.}(2020)Abdollahi, Acero, Ackermann, Ajello, Atwood,
  Axelsson, Baldini, Ballet, Barbiellini, Bastieri, Gonzalez, Bellazzini,
  Berretta, Bissaldi, Blandford, Bloom, Bonino, Bottacini, Brandt, Bregeon,
  Bruel, Buehler, Burnett, Buson, Cameron, Caputo, Caraveo, Casandjian, Castro,
  Cavazzuti, Charles, Chaty, Chen, Cheung, Chiaro, Ciprini, Cohen-Tanugi,
  Cominsky, Coronado-Bl{\'{a}}zquez, Costantin, Cuoco, Cutini, D'Ammando,
  DeKlotz, de~la Torre~Luque, de~Palma, Desai, Digel, Lalla, Mauro, Venere,
  Dom{\'{\i}}nguez, Dumora, Dirirsa, Fegan, Ferrara, Franckowiak, Fukazawa,
  Funk, Fusco, Gargano, Gasparrini, Giglietto, Giommi, Giordano, Giroletti,
  Glanzman, Green, Grenier, Griffin, Grondin, Grove, Guiriec, Harding, Hayashi,
  Hays, Hewitt, Horan, J{\'{o}}hannesson, Johnson, Kamae, Kerr, Kocevski,
  Kovac'evic', Kuss, Landriu, Larsson, Latronico, Lemoine-Goumard, Li,
  Liodakis, Longo, Loparco, Lott, Lovellette, Lubrano, Madejski, Maldera,
  Malyshev, Manfreda, Marchesini, Marcotulli, Mart{\'{\i}}-Devesa, Martin,
  Massaro, Mazziotta, McEnery, Mereu, Meyer, Michelson, Mirabal, Mizuno,
  Monzani, Morselli, Moskalenko, Negro, Nuss, Ojha, Omodei, Orienti, Orlando,
  Ormes, Palatiello, Paliya, Paneque, Pei, Pe{\~{n}}a-Herazo, Perkins, Persic,
  Pesce-Rollins, Petrosian, Petrov, Piron, Poon, Porter, Principe, Rain{\`{o}},
  Rando, Razzano, Razzaque, Reimer, Reimer, Remy, Reposeur, Romani, Parkinson,
  Schinzel, Serini, Sgr{\`{o}}, Siskind, Smith, Spandre, Spinelli, Strong,
  Suson, Tajima, Takahashi, Tak, Thayer, Thompson, Tibaldo, Torres, Torresi,
  Valverde, Klaveren, van Zyl, Wood, Yassine, \& Zaharijas}]{4fgl}
Abdollahi, S., Acero, F., Ackermann, M., {et~al.} 2020, ApJS, 247, 33,
  \dodoi{10.3847/1538-4365/ab6bcb}

\bibitem[{{Abeysekara} {et~al.}(2017){Abeysekara}, {Albert}, {Alfaro},
  {et~al.}}]{hawc}
{Abeysekara}, A.~U., {Albert}, A., {Alfaro}, R., {et~al.} 2017, ApJ, 843, 39

\bibitem[{Adrián-Martínez {et~al.}(2016)Adrián-Martínez, Ageron, Aharonian,
  {et~al.}}]{km3net}
Adrián-Martínez, S., Ageron, M., Aharonian, F., {et~al.} 2016, JPhG, 43,
  084001, \dodoi{10.1088/0954-3899/43/8/084001}

\bibitem[{Ageron {et~al.}(2011)Ageron, Aguilar, {Al Samarai}, Albert, Ameli,
  André, Anghinolfi, Anton, Anvar, Ardid, Arnaud, Aslanides, {Assis Jesus},
  Astraatmadja, Aubert, Auer, Barbarito, Baret, Basa, Bazzotti, Becherini,
  Beltramelli, Bersani, Bertin, Beurthey, Biagi, Bigongiari, Billault, Blaes,
  Bogazzi, {de Botton}, Bou-Cabo, Boudahef, Bouwhuis, Brown, Brunner, Busto,
  Caillat, Calzas, Camarena, Capone, Caponetto, Cârloganu, Carminati, Carmona,
  Carr, Carton, Cassano, Castorina, Cecchini, Ceres, Chaleil, Charvis,
  Chauchot, Chiarusi, Circella, Compère, Coniglione, Coppolani, Cosquer,
  Costantini, Cottini, Coyle, Cuneo, Curtil, D'Amato, Damy, {van Dantzig}, {De
  Bonis}, Decock, Decowski, Dekeyser, Delagnes, Desages-Ardellier, Deschamps,
  Destelle, {Di Maria}, Dinkespiler, Distefano, Dominique, Donzaud, Dornic,
  Dorosti, Drogou, Drouhin, Druillole, Durand, Durand, Eberl, Emanuele,
  Engelen, Ernenwein, Escoffier, Falchini, Favard, Fehr, Feinstein, Ferri,
  Ferry, Fiorello, Flaminio, Folger, Fritsch, Fuda, Galatá, Galeotti, Gay,
  Gensolen, Giacomelli, Gojak, Gómez-González, Goret, Graf, Guillard,
  Halladjian, Hallewell, {van Haren}, Hartmann, Heijboer, Heine, Hello, Henry,
  Hernández-Rey, Herold, Hößl, Hogenbirk, Hsu, Hubbard, Jaquet, Jaspers, {de
  Jong}, Jourde, Kadler, Kalantar-Nayestanaki, Kalekin, Kappes, Karg, Karkar,
  Karolak, Katz, Keller, Kestener, Kok, Kok, Kooijman, Kopper, Kouchner,
  Kretschmer, Kruijer, Kuch, Kulikovskiy, Lachartre, Lafoux, Lagier, Lahmann,
  Lahonde-Hamdoun, Lamare, Lambard, Languillat, Larosa, Lavalle, {Le Guen}, {Le
  Provost}, LeVanSuu, Lefèvre, Legou, Lelaizant, Lévéque, Lim, {Lo Presti},
  Loehner, Loucatos, Louis, Lucarelli, Lyashuk, Magnier, Mangano, Marcel,
  Marcelin, Margiotta, Martinez-Mora, Masullo, Mazéas, Mazure, Meli, Melissas,
  Migneco, Mongelli, Montaruli, Morganti, Moscoso, Motz, Musumeci, Naumann,
  Naumann-Godo, Neff, Niess, Nooren, Oberski, Olivetto, Palanque-Delabrouille,
  Palioselitis, Papaleo, Păvălaş, Payet, Payre, Peek, Petrovic, Piattelli,
  Picot-Clemente, Picq, Piret, Poinsignon, Popa, Pradier, Presani, Prono,
  Racca, Raia, {van Randwijk}, Real, Reed, Réthoré, Rewiersma, Riccobene,
  Richardt, Richter, Ricol, Rigaud, Roca, Roensch, Rolin, Rostovtsev, Rottura,
  Roux, Rujoiu, Ruppi, Russo, Salesa, Salomon, Sapienza, Schmitt, Schöck,
  Schuller, Schüssler, Sciliberto, Shanidze, Shirokov, Simeone, Sottoriva,
  Spies, Spona, Spurio, Steijger, Stolarczyk, Streeb, Sulak, Taiuti, Tamburini,
  Tao, Tasca, Terreni, Tezier, Toscano, Urbano, Valdy, Vallage, {Van Elewyck},
  Vannoni, Vecchi, Venekamp, Verlaat, Vernin, Virique, {de Vries}, {van Wijk},
  Wijnker, Wobbe, {de Wolf}, Yakovenko, Yepes, Zaborov, Zaccone, Zornoza, \&
  Zúñiga}]{antaresDet}
Ageron, M., Aguilar, J., {Al Samarai}, I., {et~al.} 2011, NIMPA, 656, 11,
  \dodoi{https://doi.org/10.1016/j.nima.2011.06.103}

\bibitem[{Ahlers \& Murase(2014)}]{prdkohta}
Ahlers, M., \& Murase, K. 2014, PhRvD, 90, 023010,
  \dodoi{10.1103/PhysRevD.90.023010}

\bibitem[{Albert {et~al.}(2020)Albert, R.Alfaro, Alvarez, {et~al.}}]{3hwc}
Albert, A., R.Alfaro, Alvarez, C., {et~al.} 2020, ApJ, 905, 76.
\newblock \doarXiv{2007.08582}

\bibitem[{Albert {et~al.}(2017)Albert, Andr\'e, Anghinolfi, Anton, Ardid,
  Aubert, Avgitas, Baret, Barrios-Mart\'{\i}, Basa, Belhorma, Bertin, Biagi,
  Bormuth, Bourret, Bouwhuis, Br\^anza\ifmmode~\mbox{\c{s}}\else \c{s}\fi{},
  Bruijn, Brunner, Busto, Capone, Caramete, Carr, Celli, Cherkaoui El~Moursli,
  Chiarusi, Circella, Coelho, Coleiro, Coniglione, Costantini, Coyle, Creusot,
  D\'{\i}az, Deschamps, De~Bonis, Distefano, Di~Palma, Domi, Donzaud, Dornic,
  Drouhin, Eberl, El~Bojaddaini, El~Khayati, Els\"asser, Enzenh\"ofer,
  Ettahiri, Fassi, Felis, Fusco, Galat\`a, Gay, Giordano, Glotin, Gr\'egoire,
  Gracia~Ruiz, Graf, Hallmann, van Haren, Heijboer, Hello, Hern\'andez-Rey,
  H\"o\ss{}l, Hofest\"adt, Hugon, Illuminati, James, de~Jong, Jongen, Kadler,
  Kalekin, Katz, Kie\ss{}ling, Kouchner, Kreter, Kreykenbohm, Kulikovskiy,
  Lachaud, Lahmann, Lef\`evre, Leonora, Lotze, Loucatos, Marcelin, Margiotta,
  Marinelli, Mart\'{\i}nez-Mora, Mele, Melis, Michael, Migliozzi, Moussa,
  Navas, Nezri, Organokov, P\ifmmode \u{a}\else \u{a}\fi{}v\ifmmode \u{a}\else
  \u{a}\fi{}la\ifmmode~\mbox{\c{s}}\else \c{s}\fi{}, Pellegrino, Perrina,
  Piattelli, Popa, Pradier, Quinn, Racca, Riccobene, S\'anchez-Losa, Salda\~na,
  Salvadori, Samtleben, Sanguineti, Sapienza, Sch\"ussler, Sieger, Spurio,
  Stolarczyk, Taiuti, Tayalati, Trovato, Turpin, T\"onnis, Vallage,
  Van~Elewyck, Versari, Vivolo, Vizzoca, Wilms, Zornoza, \&
  Z\'u\~niga}]{antaresDet2}
Albert, A., Andr\'e, M., Anghinolfi, M., {et~al.} 2017, PhRvD, 96, 082001,
  \dodoi{10.1103/PhysRevD.96.082001}

\bibitem[{Albert {et~al.}(2018)Albert, Andr{\'{e}}, Anghinolfi, Anton, Ardid,
  Aubert, Aublin, Avgitas, Baret, Barrios-Mart{\'{\i}}, Basa, Belhorma, Bertin,
  Biagi, Bormuth, Boumaaza, Bourret, Bouwhuis, Br{\^{a}}nza{\c{s}}, Bruijn,
  Brunner, Busto, Capone, Caramete, Carr, Celli, Chabab, Moursli, Chiarusi,
  Circella, Coelho, Coleiro, Colomer, Coniglione, Costantini, Coyle, Creusot,
  D{\'{\i}}az, Deschamps, Distefano, Palma, Domi, Don{\`{a}}, Donzaud, Dornic,
  Drouhin, Eberl, Bojaddaini, Khayati, Elsässer, Enzenhöfer, Ettahiri, Fassi,
  Felis, Fermani, Ferrara, Fusco, Gay, Glotin, Gr{\'{e}}goire, Ruiz, Graf,
  Hallmann, van Haren, Heijboer, Hello, Hern{\'{a}}ndez-Rey, Hö{\ss}l,
  Hofestädt, Illuminati, de~Jong, Jongen, Kadler, Kalekin, Katz,
  Khan-Chowdhury, Kouchner, Kreter, Kreykenbohm, Kulikovskiy, Lachaud, Lahmann,
  Lef{\`{e}}vre, Leonora, Lotze, Loucatos, Marcelin, Margiotta, Marinelli,
  Mart{\'{\i}}nez-Mora, Mele, Melis, Migliozzi, Moussa, Navas, Nezri,
  Nu{\~{n}}ez, Organokov, P{\u{a}}v{\u{a}}la{\c{s}}, Pellegrino, Piattelli,
  Popa, Pradier, Quinn, Racca, Randazzo, Riccobene, S{\'{a}}nchez-Losa,
  Salda{\~{n}}a, Salvadori, Samtleben, Sanguineti, Sapienza, Schüssler,
  Spurio, Stolarczyk, Taiuti, Tayalati, Trovato, Vallage, Elewyck, Versari,
  Vivolo, Wilms, Zaborov, Zornoza, \& and}]{antaresPublic}
Albert, A., Andr{\'{e}}, M., Anghinolfi, M., {et~al.} 2018, ApJ, 863, L30,
  \dodoi{10.3847/2041-8213/aad8c0}

\bibitem[{Albert {et~al.}(2021)Albert, Alvarez, Camacho,
  Arteaga-Vel{\'{a}}zquez, Arunbabu, Rojas, Solares, Baghmanyan,
  Belmont-Moreno, BenZvi, Brisbois, Caballero-Mora, Capistr{\'{a}}n,
  Carrami{\~{n}}ana, Casanova, Cotti, Cotzomi, de~Le{\'{o}}n, la~Fuente,
  Dingus, DuVernois, Durocher, D{\'{\i}}az-V{\'{e}}lez, Engel, Espinoza, Fan,
  Alonso, Fleischhack, Fraija, Galv{\'{a}}n-G{\'{a}}mez, Garc{\'{\i}}a,
  Garc{\'{\i}}a-Gonz{\'{a}}lez, Garfias, Gonz{\'{a}}lez, Goodman, Harding,
  Hern{\'{a}}ndez, Hona, Huang, Hueyotl-Zahuantitla, Hüntemeyer, Iriarte,
  Jardin-Blicq, Joshi, Kieda, Kunde, Lara, Lee, Vargas, Linnemann, Longinotti,
  Luis-Raya, Lundeen, Malone, Mart{\'{\i}}nez, Martinez-Castellanos,
  Mart{\'{\i}}nez-Castro, Matthews, Miranda-Romagnoli, Morales-Soto, Moreno,
  Mostaf{\'{a}}, Nayerhoda, Nellen, Newbold, Nisa, Noriega-Papaqui,
  Olivera-Nieto, Peisker, P{\'{e}}rez-P{\'{e}}rez, Rho, Rosa-Gonz{\'{a}}lez,
  Ruiz-Velasco, Salazar, Greus, Sandoval, Schneider, Schoorlemmer, Smith,
  Springer, Tollefson, Torres, Torres-Escobedo, Ure{\~{n}}a-Mena,
  Villase{\~{n}}or, Weisgarber, Willox, Zepeda, Zhou, \&
  de~Le{\'{o}}n~and}]{hawc_agn}
Albert, A., Alvarez, C., Camacho, J. R.~A., {et~al.} 2021, ApJ, 907, 67,
  \dodoi{10.3847/1538-4357/abca9a}

\bibitem[{{Ayala Solares} {et~al.}(2021)}]{amonHAWCIC}
{Ayala Solares}, {et~al.} 2021, ApJ, 906, 63, \dodoi{10.3847/1538-4357/abcaa4}

\bibitem[{Ayala~Solares {et~al.}(2019)Ayala~Solares, Cowen, DeLaunay, Fox,
  Keivani, Mostafá, Murase, Turley, Albert, André, \&
  et~al.}]{amonfermi_antares}
Ayala~Solares, H.~A., Cowen, D.~F., DeLaunay, J.~J., {et~al.} 2019, ApJ, 886,
  98, \dodoi{10.3847/1538-4357/ab4a74}

\bibitem[{{Ayala Solares} {et~al.}(2020){Ayala Solares}, Coutu, Cowen,
  DeLaunay, Fox, Keivani, Mostaf{\'a}, Murase, Oikonomou, Seglar-Arroyo, Te{\v
  s}i{\'c}, \& Turley}]{amon2020}
{Ayala Solares}, H.~A., Coutu, S., Cowen, D., {et~al.} 2020, APh, 114, 68 ,
  \dodoi{https://doi.org/10.1016/j.astropartphys.2019.06.007}

\bibitem[{Barthelmy(1990)}]{gcn}
Barthelmy, S. 1990, Galactic Coordinates website,  NASA.
\newblock \url{https://gcn.gsfc.nasa.gov/}

\bibitem[{{Brady} {et~al.}(2019)}]{scimma2}
{Brady}, P., {et~al.} 2019, Scalable Cyberinfrastructure to support
  Multi-Messenger Astrophysics.
\newblock \url{https://scimma.org/index.html}

\bibitem[{Chang {et~al.}(2019)}]{scimma}
Chang, P., {et~al.} 2019, arXiv.
\newblock \doarXiv{1903.04590}

\bibitem[{Cowan(2002)}]{analysisbook}
Cowan, G. 2002, Statistical Data Analysis (Oxford University Press)

\bibitem[{{Dom{\'\i}nguez} {et~al.}(2011){Dom{\'\i}nguez}, {Primack},
  {Rosario}, {et~al.}}]{eblDominguez}
{Dom{\'\i}nguez}, A., {Primack}, J.~R., {Rosario}, D.~J., {et~al.} 2011, MNRAS,
  410, 2556, \dodoi{10.1111/j.1365-2966.2010.17631.x}

\bibitem[{Fisher(1938)}]{fisher}
Fisher, R.~A. 1938, Statistical methods for research workers (Edinburgh, Oliver
  and Boyd)

\bibitem[{{Hunter}(2007)}]{matplotlib}
{Hunter}, J. 2007, CSE, 9, 90

\bibitem[{{Illuminati} {et~al.}(2019){Illuminati}, {Aublin}, \&
  {Navas}}]{giuliaICRC}
{Illuminati}, G., {Aublin}, J., \& {Navas}, S. 2019, in International Cosmic
  Ray Conference, Vol.~36, 36th International Cosmic Ray Conference (ICRC2019),
  920.
\newblock \doarXiv{1908.08248}

\bibitem[{{M}c{K}inney(2010)}]{pandas}
{M}c{K}inney, W. 2010, in {P}roceedings of the 9th {P}ython in {S}cience
  {C}onference, ed. {S}t\'efan van~der {W}alt \& {J}arrod {M}illman, 56 -- 61,
  \dodoi{10.25080/Majora-92bf1922-00a}

\bibitem[{Murase \& Bartos(2019)}]{mmsources}
Murase, K., \& Bartos, I. 2019, ARNPS, 69, 477,
  \dodoi{10.1146/annurev-nucl-101918-023510}

\bibitem[{{Price-Whelan} {et~al.}(2018){Price-Whelan}, {Sip{\H{o}}cz},
  {G{\"u}nther}, {et~al.}}]{astropy}
{Price-Whelan}, A.~M., {Sip{\H{o}}cz}, B.~M., {G{\"u}nther}, H.~M., {et~al.}
  2018, AJ, 156, 123, \dodoi{10.3847/1538-3881/aabc4f}

\bibitem[{{Strolger} {et~al.}(2015){Strolger}, {Dahlen}, {Rodney},
  {et~al.}}]{ccsnr2015}
{Strolger}, L.-G., {Dahlen}, T., {Rodney}, S.~A., {et~al.} 2015, ApJ, 813, 93,
  \dodoi{10.1088/0004-637X/813/2/93}

\bibitem[{Taboada {et~al.}(2017)Taboada, Tung, \& Wood}]{firesong}
Taboada, I., Tung, C.~F., \& Wood, J. 2017, in "Proceedings of 35th
  International Cosmic Ray Conference {\textemdash} PoS(ICRC2017)", Vol. 301,
  663, \dodoi{10.22323/1.301.0663}

\bibitem[{Turley {et~al.}(2018)Turley, Fox, Keivani, DeLaunay, Cowen,
  Mostaf{\'{a}}, Solares, \& Murase}]{amonfermi_icecube}
Turley, C.~F., Fox, D.~B., Keivani, A., {et~al.} 2018, ApJ, 863, 64,
  \dodoi{10.3847/1538-4357/aad195}

\bibitem[{{Van der Walt} {et~al.}(2011){Van der Walt}, {Colbert}, \&
  {Varoquaux}}]{numpy}
{Van der Walt}, S., {Colbert}, S.~C., \& {Varoquaux}, G. 2011, CSE, 13, 22

\bibitem[{{Virtanen} {et~al.}(2020){Virtanen}, {Gommers}, {Oliphant},
  {et~al.}}]{scipy}
{Virtanen}, P., {Gommers}, R., {Oliphant}, T.~E., {et~al.} 2020, Nat Methods,
  17, 261, \dodoi{https://doi.org/10.1038/s41592-019-0686-2}

\bibitem[{{Wenger} {et~al.}(2000){Wenger}, {Ochsenbein}, {Egret},
  {et~al.}}]{simbad}
{Wenger}, M., {Ochsenbein}, F., {Egret}, D., {et~al.} 2000, AAPS, 143, 9,
  \dodoi{10.1051/aas:2000332}

\end{thebibliography}
